# Locating Power System Oscillation Sources by Extracting Interharmonics from Synchrophasor Data

Sophia Xu, Student M.IEEE
*Dept. of Computer Science*
University of British Columbia
Vancouver, Canada

Wilsun Xu, F.IEEE
*Dept. of Eletrical & Computer Engineering*
University of Alberta
Edmonton, Canada

***Abstract* — Recent studies have shown that oscillating phasors arise from beating waves caused by interharmonics. This finding has led to an interharmonic-based oscillation source-location method. But waveform data needed for the method are less available than PMU synchrophasor data. This paper investigates whether interharmonics can be extracted directly from phasor data for similar applications. The results show that, for certain oscillation phenomena, phasor data can support interharmonic extraction and oscillation source location. Sensitivity studies have identified the requirements, and a cloud-based software tool is developed for PMU-based source location.**

## I. Introduction

Power system oscillations have long concerned system operators [1]. Although power system stabilizers helped address synchronous-generator-driven oscillations after their introduction in the 1980s, the problem has re-emerged with the growing interconnection of inverter-based resources (IBRs) [2]. IBRs can introduce oscillations over a wide frequency range, from a few hertz to tens of hertz, and PMU-based wide-area monitoring systems have made these events more visible.

Among the major research needs in this area, oscillation source location is particularly important because identifying the source is a prerequisite for mitigation. Reference [3] presents a good review of the state of this research. However, almost most methods rely on heuristic formulations of the source location problem such as using Lyapunov energy functions. Their validity is difficult to prove beyond case studies [4].

A recent work on waveform-based analysis of oscillating phasors has brought new opportunities [5]. This research shows that phasor oscillations are manifestations of beating waves caused by interharmonic spectral components. Interharmonics carry oscillatory energy and propagate through the grid, producing system-wide oscillations. Based on this new insight, an energy conservation law based, mathematically proven method has been developed to locate oscillation sources as well as identify resonant components successfully.

These methods require waveform data, which are not widely available. On the other hand, PMU-measured phasor data are more available through the wide-area monitoring systems. This raises an important question: can interharmonic components be extracted from phasor data to support similar applications without waveform measurements? This paper presents findings on this question and shows that, for certain oscillation phenomena, interharmonics can be extracted from phasor data.

The remainder of this paper is organized as follows. Section II reviews the interharmonic interpretation of power system oscillations. Section III presents the proposed method and measurement infrastructure. Section IV provides verification and sensitivity study results. Section V describes a cloud-based software implementation intended to support researchers and engineers to analyze field measured PMU data.

## II. Technical Background

Oscillations have traditionally been analyzed using phasor data, such as those recorded by PMUs. Examining the waveforms underlying the phasors through spectral analysis, however, reveal new understanding into the mechanisms that drive these oscillations. This section summarizes the key findings from waveform-based oscillation analysis.

### *A. Beating Wave Phnomenon Unlying Oscillating Phasor*

Fig.1 shows a measured current waveform and its spectrum of a wind farm during an oscillation event. (The fundamental frequency component is removed in the spectral plot.) A wave-beating pattern can be seen clearly. The corresponding phasor is seen as an oscillating red curve. Fourier analysis of the waveform reveals a large spectral component at 63.5 Hz and a small one at 56.5Hz. According to IEC 61000-4-30 [6], these components are called interharmonics, which are defined as the spectral components reside between harmonic frequencies. As such, interharmonics are also called non-integer harmonics.

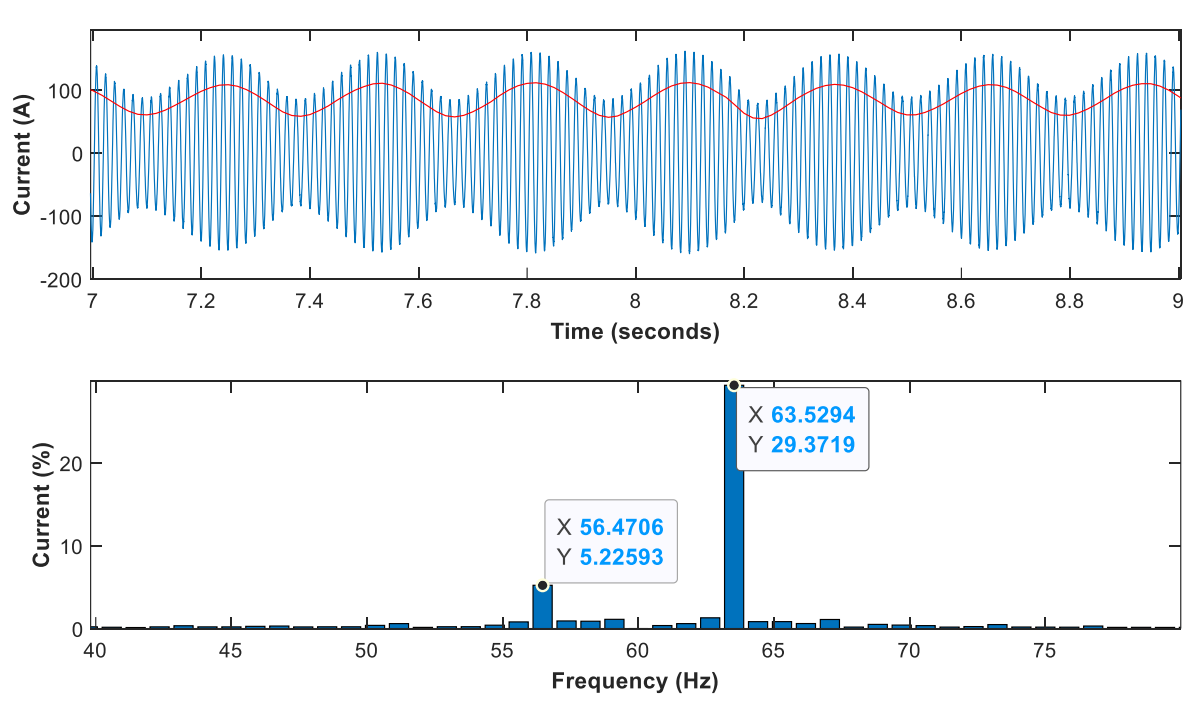

Fig.1. Sample current of a windfarm.

The mechanism for interharmonics to cause phasor oscillation can be understood from Fig.2. This figure depicts a waveform $v(t)$ that is composed of a fundamental frequency component $v_1(t)$ and one interharmonic component $v_{IH}(t)$. The magnitude of resulting phasor is also shown and labeled as $V_{phasor}$, which represents essentially the RMS value of $v(t)$ in each window or $v_1(t)$ cycle.

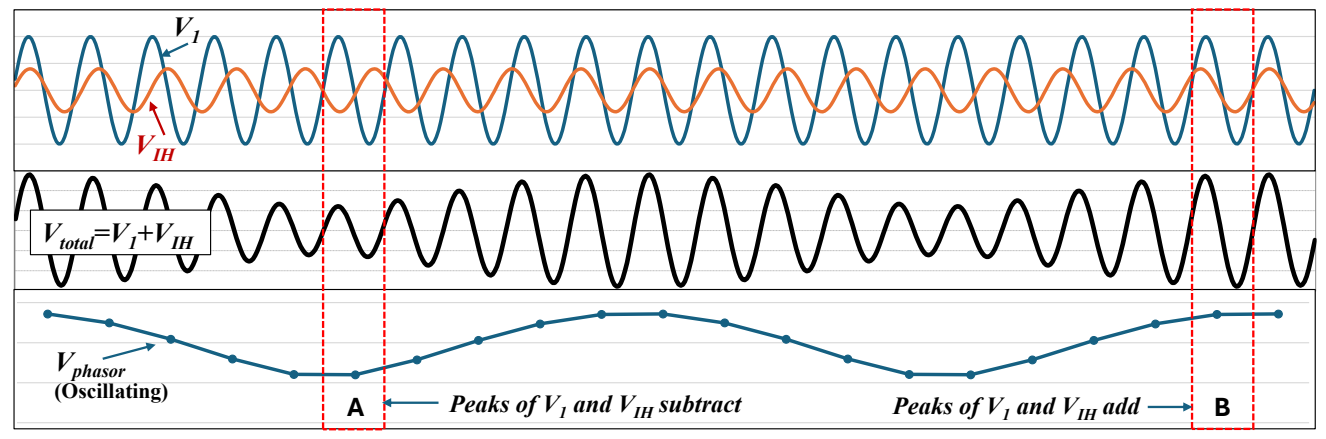

Fig. 2. Mechanism of an interharmonic causing phasor oscillation (©IEEE).

As shown in Fig. 2, the peaks of $v_{ih}(t)$ are not synchronized with those of $v_1(t)$. Consequently, the two components add together at cycle B and subtract at cycle A. The phasor magnitude at cycle B is therefore greater than that at cycle A, creating the appearance of an oscillation in the phasor magnitude. This is the wave-beating phenomenon produced by the interaction of two waves with different frequencies, and *the beating frequency is the phasor oscillation frequency*.

Reference [6] has established that the presence of interharmonics is a necessary and sufficient condition for a phasor to exhibit oscillatory behavior. This statement should not be interpreted as implying that interharmonics are always the root cause of an oscillation event. Rather, *interharmonics act as carriers of oscillatory energy*, enabling the energy to propagate in a grid and giving rise to system-wide oscillations.

### B. New Insights on Power System Oscillations

The insights gained from the interharmonic-interpretation of power system oscillations that are relevant to this work are summarized below:

Sources of Interharmonics: A grid has many natural frequencies. They can be described using the imaginary parts of the eigenvalues of a linearized circuit model. If a grid has near zero damping at one particular natural frequency, "ringing" at that frequency will sustain. *This frequency is the interharmonic frequency* that beats with the regular fundamental frequency in the grid, causing apparent phasor oscillation. The "ringing" or interharmonic is caused by interharmonic energy producing components, such as equivalent negative resistances in the case of IBRs [7], interacting with the RLC components in the grid.

Oscillation frequency versus interharmonic frequency: An oscillation event involves at least two distinct frequencies: the frequency of phasor oscillation $f_{os}$ and the frequency of the interharmonic causing phasor oscillation $f_{ih}$. The two frequencies are related to each through the following formulas:

$$f_{beat} = f_{os} = | f_{ih} - Round(\frac{f_{ih}}{f_1}) f_1 | \quad (1)$$

$$f_{ih} = | hf_1 \pm f_{os} | \qquad h = 0,1,2,3.. \quad (2)$$

where Round*(*)* means rounding to the nearest integer. The above formula reveals the following: 1) the highest $f_{os}$ reportable by a phasor is 30Hz for a 60Hz system, and 2) different $f_{ih}$ can yield the same $f_{os}$. For example, interharmonics at 15 Hz, 45 Hz, and 75 Hz all produce an oscillation frequency of 15 Hz. This is caused by the aliasing effect as PMUs attempt to represent one cycle of waveform data using one phasor data, i.e. it has a 60Hz "sample" rate.

Modulated Wave versus Beating Wave: A common view in power system analysis is that phasor oscillations result from sine-wave modulation. Cyclic loads can indeed produce waveform behavior that appears modulation-like; however, closer examination still reveals the existence of interharmonic components. The following is a magnitude-modulated phasor:

$$\vec{I} = m(t)\angle\theta, \qquad m(t) = I_0 + \Delta I \cos(\omega_{os} t)$$

The actual waveform of this phasor is

$$\begin{aligned} i(t) &= m(t)\cos(\omega_1 t + \theta) = [I_o + \Delta I \cos(\omega_{os} t)]\cos(\omega_1 t + \theta) \\ &= I_o \cos(\omega_1 t + \theta) + \Delta I[\cos(\omega_{os} t)\cos(\omega_1 t + \theta)] \\ &= I_o \cos(\omega_1 t + \theta) + \Delta I \cos[(\omega_1 + \omega_{os})t + \theta] + \Delta I \cos[(\omega_1 - \omega_{os})t + \theta] \end{aligned}$$

Two interharmonics with frequencies $\omega_1 \pm \omega_{os}$ are observed. The key findings are as follows:

- A magnitude-modulated phasor corresponds to a specific wave-beating pattern characterized by two interharmonics with equal magnitudes and phase angles, located symmetrically around $f_1$.
- If a waveform contains two interharmonics with different magnitudes as in Fig. 1, a modulated phasor cannot properly represent the beating pattern or the true oscillatory behavior.
- Small-disturbance stability-related natural oscillations typically have only one sustained ringing frequency; therefore, a modulated phasor is not a valid model for such oscillatory behavior.

Modulated phasors have worked well for synchronous-generator oscillations because their oscillation frequencies are typically below 2Hz. In that range, the terms $\cos[(\omega_1 \pm \omega_{os}) + \theta]$ remain nearly equal over short intervals. For higher $f_{os}$ events, especially IBR-related oscillations above 10 Hz, the underlying interharmonics can differ substantially from 60 Hz, making the modulated-phasor model unreliable.

### C. Oscillation Source Location Using Interharmonic Power

The interharmonic insights lead to a straightforward method for oscillation source location: Phasor oscillations indicate the presence of interharmonics in the underlying waveforms, and power at the interharmonic frequencies is required to drive the flow of interharmonics through the system. Therefore, facilities that produce interharmonic power can be identified as oscillation sources. Comparing the interharmonic power contributions from different facilities one can locate multiple sources and rank their contributions.

Using four types of field measurements - synchronous generators, solar farms, wind farms, and loads with variable-frequency drives - [5] demonstrated that the interharmonic power method can successfully locate and rank oscillation sources. The study also showed that interharmonic power represents the sensitivity of the critical eigenvalue to the equivalent resistance of the monitored facility.

## III. Using Phasor Data for Source Location

The interharmonic power method should be implemented using waveform data. Waveform data contains full information about the interharmonics involved in an oscillation event. If waveform data is not available, is it possible to extract interharmonics from the synchrophasor data collected by the PMUs? A method is proposed in this section to achieve the goal.

### A. Problem definition

The oscillation source location problem is to identify the generating or load facilities responsible for an oscillation event using only synchrophasor data measured at facility–grid interconnection points, such as points a–e in Fig. 3. This monitoring infrastructure is already widely available because of interconnection requirements such as IEEE Std. 2800 [8]. PMUs may also be installed on interties and major transmission lines, as indicated by points 1–3; however, those measurements are not required for source identification because interconnection-point measurements capture much more facility specific information relevant to the behavior of the facility.

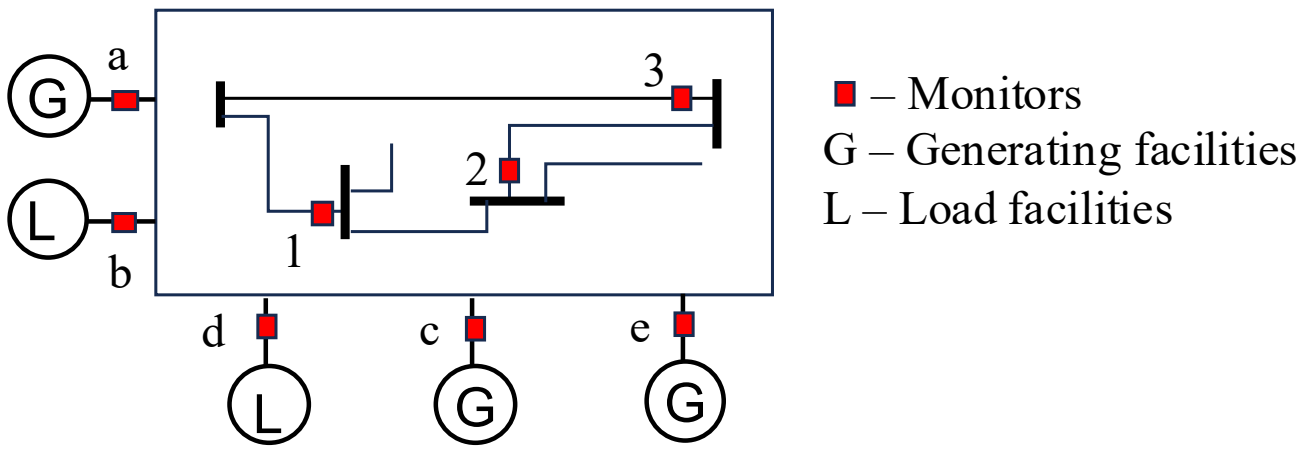

Fig.3. Infrastructure for oscillation monitoring.

Per PMU measurement standard [9], the k-th cycle phasor are determined using the following operations:

$$\vec{V}_{phasor}(t_k) = \frac{\sqrt{2}}{G}\sum_{i=0}^{N-1} w(i+1)v(i\Delta T + t_{start})e^{-j\omega_0(i\Delta T + t_{start})}$$
$$G = \sum_{i=1}^{N} w(i) \tag{3}$$

where

- $t_k$ is the time stamp assigned to the k[th] phasor output,
- $t_{start}$ is the precision time of the first waveform sample used for phasor calculation,
- N is the number of samples used for phasor calculation. N covers one fundamental frequency (60Hz) cycle.
- $w(i)$ is a weighting function;
- $\Delta T = T_0/N$ is the sampling interval, and $T_0=(1/f_0)$ is the period of one fundamental frequency cycle (60Hz or 50Hz), and $\omega_o=2\pi f_0$.

It is useful to note that a PMU may output one phasor data per one cycle, per two cycle etc. $f_0$ is a constant (60Hz or 50Hz) and it may not correspond to actual grid frequency of $f_1$.

### B. The Proposed Method

The first step is to estimate the oscillation frequency $f_{os}$ based on the oscillation pattern of the phasor magnitude. Various methods can be used to determine the frequency. Our studies show that a simple FFT method that uses phasor magnitude data covering multiple oscillation period as input is sufficient to estimate $f_{os}$ adequately. This is fully supported by sensitivity analysis presented in Section IV.B.

Based on the beat frequency formula Eq.(2), we can assume that the following two interharmonic components might exist in the waveform:

$$f_{IH1} = f_1 - f_{os} \qquad f_{IH2} = f_1 + f_{os} \tag{4}$$

The above assumption is based on the common interharmonic frequency range of current power systems. It is valid for oscillations involving synchronous generators, wind farms, solar farms and cyclic loads. It may not be valid for variable frequency drives and SSR. The proposed method allows one to make a different assumption for the frequencies such as $f_{IH1}=f_{os}$, $f_{IH2}=f_1-f_{os}$ based on engineering judgement of the case. The need to estimate interharmonic frequencies properly is one deficiency of using the PMU data. This is caused by aliasing effect inherent to phasor data as explained in Section II.

Once the interharmonic frequencies are "known", the waveform can be constructed by assuming they contain only the fundamental and interharmonic components as follows:

$$\begin{aligned} v(t) &== \sqrt{2}V_1\cos(\omega_1 t+\delta) + \sqrt{2}V_2\cos(\omega_2 t+\delta_2) + \sqrt{2}V_3\cos(\omega_3 t+\delta_3) \\ &= \sqrt{2}[X_1\cos(\omega_1 t) - Y_1\sin(\omega_1 t) \\ &+ X_2\cos(\omega_2 t) - Y_2\sin(\omega_2 t) + X_3\cos(\omega_3 t) - Y_3\sin(\omega_3 t)] \end{aligned} \tag{5}$$

where $\omega_1=2\pi f_1$ , $\omega_2=2\pi f_{IH1}$, and $\omega_3=2\pi f_{IH2}$. Coefficients $X_1$~$Y_3$ are unknowns to be solved. Using the above waveform model and Eq.(3), the phasor data for the k-th cycle becomes:

$$\begin{aligned} \vec{V}_{phasor}(t_k) &= X_1 W_{1\cos-k} + Y_1 W_{1\sin-k} \\ &+ X_2 W_{2\cos-k} + Y_2 W_{2\sin-k} + X_3 W_{3\cos-k} + Y_3 W_{3\sin-k} \end{aligned} \tag{6}$$

$W_{1cos-k}$ etc. are known coefficients derived from Eq. (3) and (5). One example W coefficient is shown below:

$$W_{1\cos-k} = \frac{2}{G}\sum_{i=0}^{N-1} w(i+1)\cos[\omega_1(i\Delta T + t_{start})]e^{-j\omega_0(i\Delta T + t_{start})} \tag{7}$$

If there are m cycles of phasor data in one period of oscillation (k=1...m), there will be m equations similar to the one shown above with different $V_{phasor}(t_k)$ and $W$ coefficient values. Thus, six unknowns $X_1$~$Y_3$ can be solved using least square fitting to the m phasor data.

Since each phasor data is a complex number and there are 6 real-numbered unknowns, three phasor data is sufficient in theory to solve for the unknowns. In order to have adequate redundancy, the number of phasor data required for reliable solution should be at least 6. They should ideally cover one or two whole oscillation periods.

### C. Implementation Steps

The full method can be summarized as follows:

1) Take in the phasor data of the oscillation event to be processed;

2) Estimate $f_{os}$ using the oscillation-containing segment of the data and a standard FFT operation;
3) Estimate the interharmonic frequencies $f_{IH1}$ and $f_{IH2}$ based on Eq.(2) and engineering knowledge;
4) Step through each oscillation period as follows:
   - Compute $W$ coefficients for the period
   - Solve $X_i$ and $Y_i$ using least square fitting
   - Calculate interharmonic powers using the results
5) Check the signs of interharmonic powers against that of the fundamental frequency power. If the facility generates power, same signs of $P_1$ and $P_{IH}$ means the facility is a source. If it is a load facility, opposite sign means source.
6) If needed, compare the interharmonic powers of different facilities. They quantify the level of contributions of the facilities.

If one of the assumed interharmonic components is not actually present in the waveform, the estimated X and Y coefficients of that component will be very small. It is important to note that if one assumes there are more than two interharmonic components based on Eq.(2), the least square method will encounter ill-conditioned matrix since the phasor data don't contain enough information to solve such a case.

## IV. Verification and Sensitivity Studies

### A. Verification Study

The proposed method has been tested using multiple field-data sets. Representative cases are presented here. Fig. 4 shows the wind farm case of Fig. 1. Phasor data were first computed from the waveform using the PMU algorithm to emulate PMU measurements, then used as input to the proposed algorithm to estimate interharmonic voltages, currents, and powers. The top chart compares interharmonic currents extracted from waveform and phasor data, while the bottom chart compares three-phase interharmonic powers. The close agreement indicates that phasor data can provide acceptable source location performance.

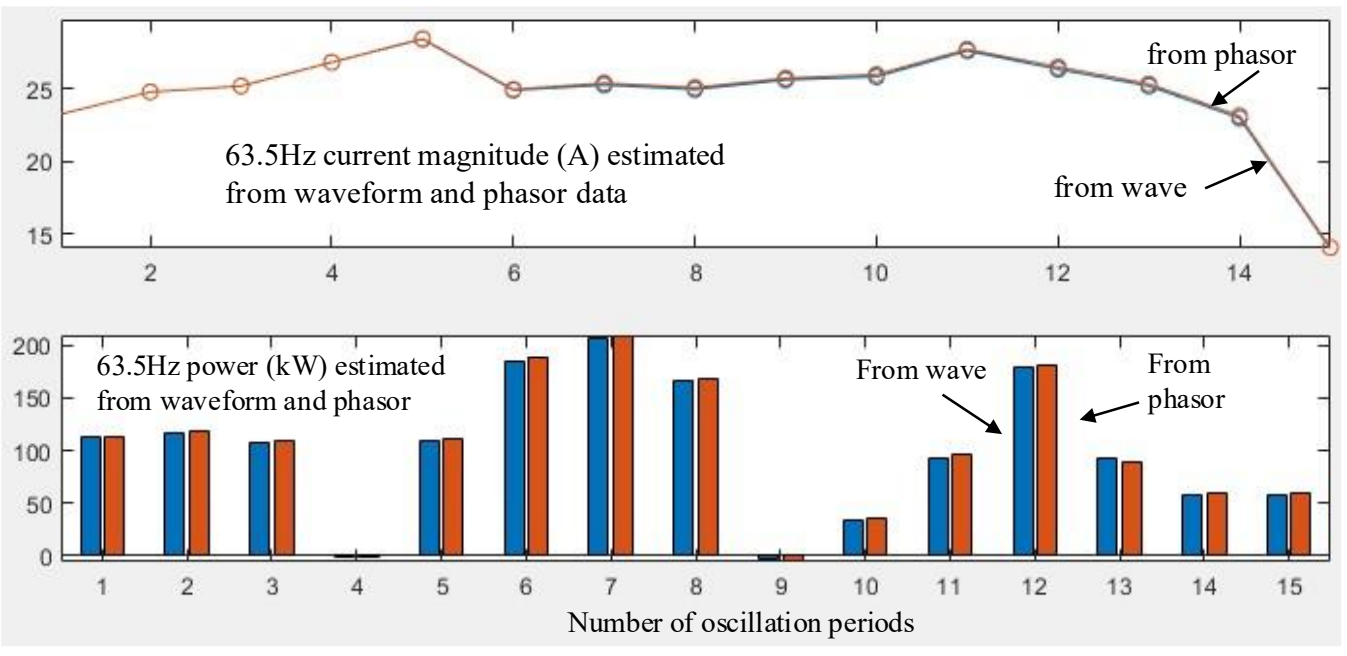

Fig. 4. Comparison of wind farm results.

The case of a solar farm is shown in Fig. 5. The main interharmonic frequency is 75Hz which gives an oscillation frequency of 15Hz. The comparison of the interharmonic estimation results is shown in Fig. 6. The differences between waveform- and phasor-based results are more evident because the solar-farm current has a higher interharmonic frequency and exhibits stronger beating patterns. Although phasor accuracy is reduced, the results remain practically acceptable when PMU data are the only available measurements.

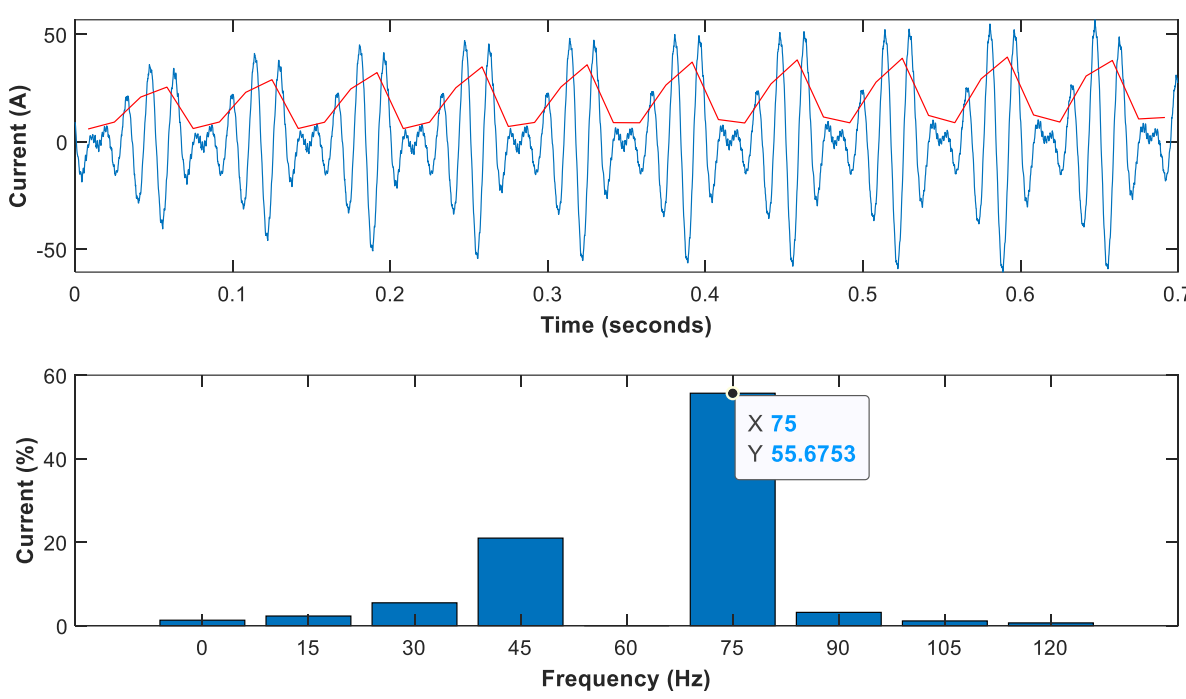

Fig. 5. Current of a solar farm.

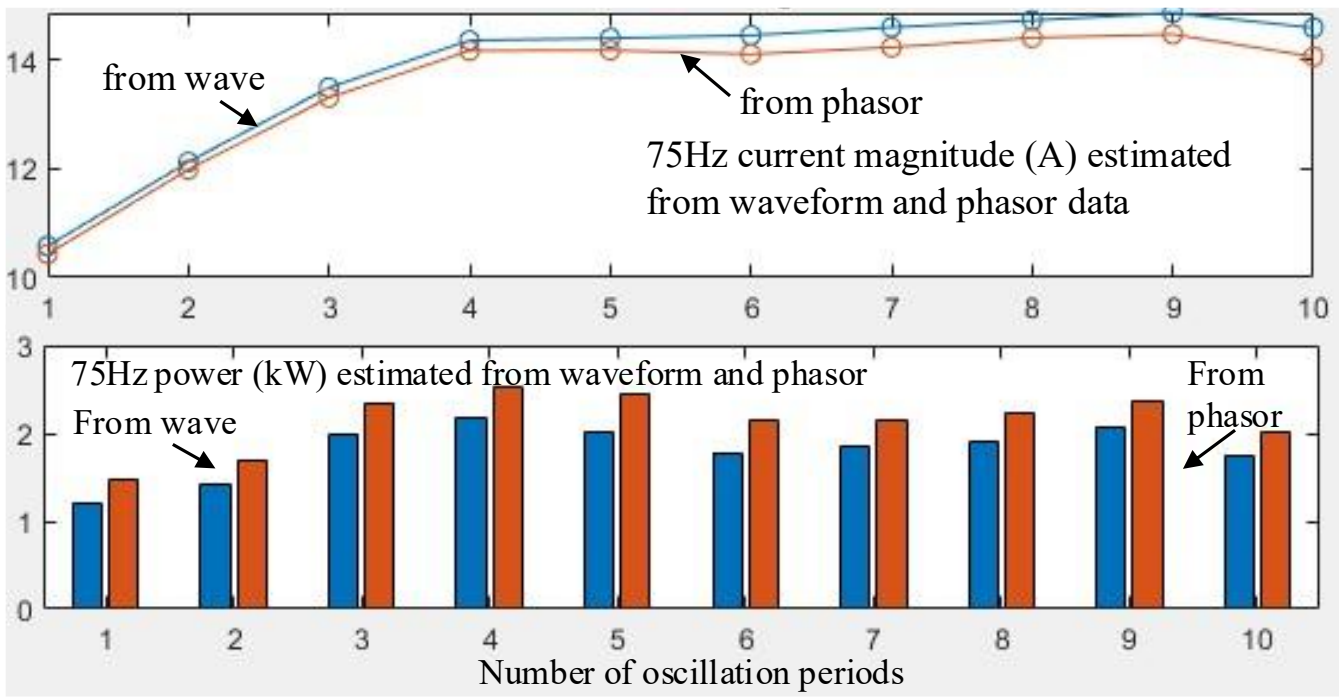

Fig. 6. Comparison of solar farm results.

For both wind and solar farms cases, the interharmonic powers are positive, indicating they are sources of oscillation. This conclusion has been confirmed by field people and other troubleshooting analysis performed by the utilities [10].

### B. Sensitivty Study

In practice, some PMU implementation details may not be available to users. Examples are sampling rate N, weighting factors *w(i)* and so on. It is therefore important to identify which input parameters must be known for reliable estimation. The following are the main findings. Due to space limitation, only shown below are the results related to IH2 interharmonic which is the interharmonic of interest for source location:

- <u>PMU internal sampling rate:</u> This is related to parameter ΔT or N in Eq.(3) and (7) which may be unknown to the users. Top chart of Fig. 7 shows the case of assuming N as 4, 16, 32 and 64 while the actual N is 128 samples/cycle. It is found that underestimating N only causes noticeable error when the value is too low. Bottom chart of Fig. 7 shows the case of overestimating N, i.e. when actual N=4 but the assumed N is 16, 32 ... The error rises once the assumed N exceeds the actual rate but then plateaus around 5–6%. This suggests overestimation carries a limited impact, whereas underestimation may cause more errors. As a result, assuming a higher sampling rate such as N=16 or 32 is a safer bet if actual N value is unknown.

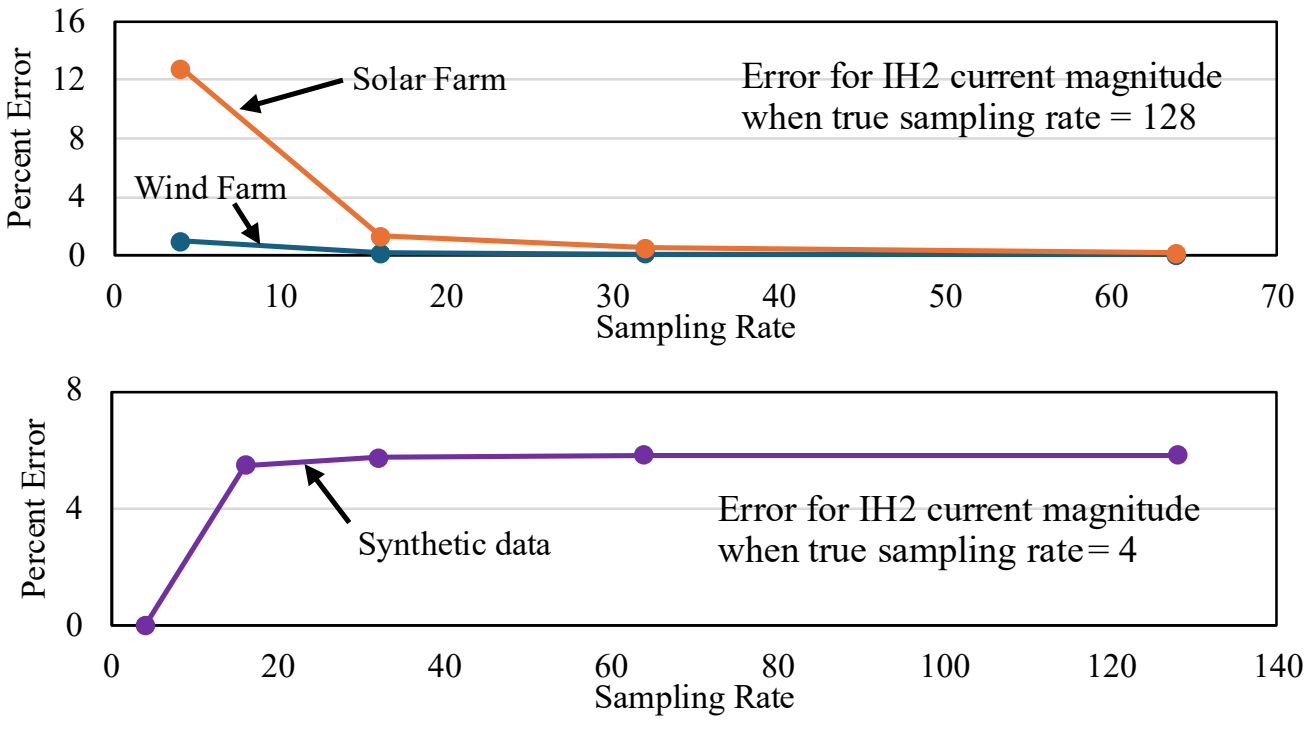

Fig. 7. Impact of assumed PMU sampling rate.

- PMU timestamp: This is about $t_k$ which is related to $T_{start}$ in Eqs. (3) and (7). Since PMU compresses one cycle of data from $T_{start}$ to $T_{start}+(N-1)\Delta T$ into one data point, $t_k$ could be assigned to any time inside this period. The PMU standard recommends $t_k$ be assigned to the middle point time, i.e. $t_k=T_{start}+0.5(N-1)\Delta T$. Fig. 8 shows the impact of $t_k$ as a function of m defined as $t_k=T_{start}+m(N-1)\Delta T$. It can be seen that m has a large impact on the accuracy. Therefore, one must use the data with a timestamp corresponding to that used by the PMU. If no such information is available, use m=50% which means the phasor data follows standard.

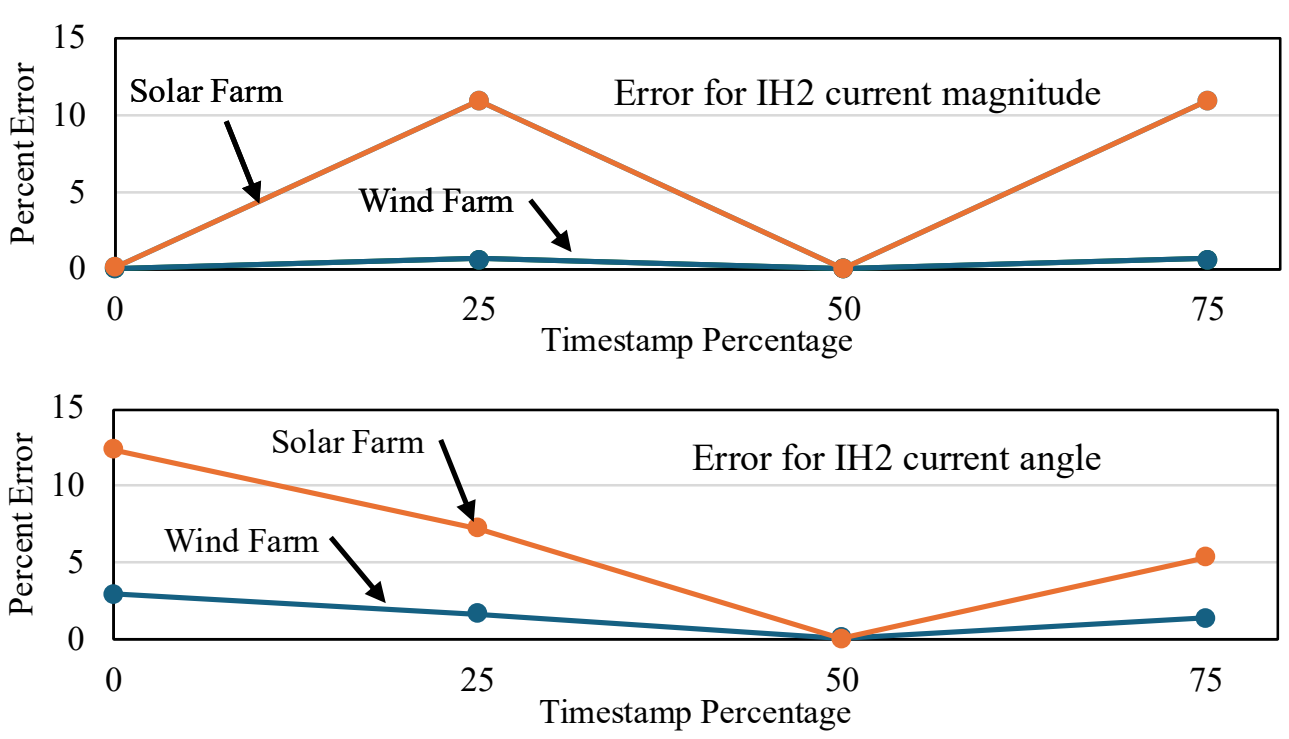

Fig. 8. Impact of assumed PMU timestamp.

- PMU weighting function: This is related to function w(t) in Eqs. (3) and (7). Sensitivity studies using simulated waveforms show that the weighting function does not affect the estimation results significantly. The simulated waveform has the following form:

$$v(t) == \sqrt{2}V_1 \cos(\omega_1 t+\delta) + \sqrt{2}V_2 \cos(\omega_2 t+\delta_2)+\sqrt{2}V_3 \cos(\omega_3 t+\delta_3) \quad (8)$$

The impact of interharmonic frequency on the estimation accuracy is also investigated. Since the highest reportable oscillation frequency by a PMU is 30Hz for a 60Hz system (Section II), it is reasonable to expect that the estimation error will increase if interharmonic frequency increases. Using synthetic waveform of Eq.(8) as study case, Fig. 9 confirms this prediction. It shows that when interharmonic frequency is above 80Hz, the error starts to increase. This finding implies that if $f_{os}$>20Hz, PMU phasor data cannot be used for interharmonic extraction or source location.

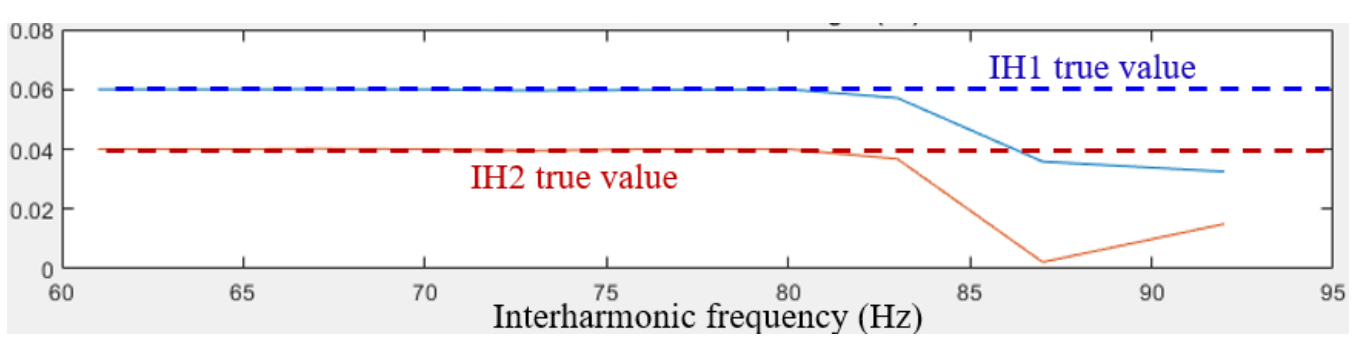

Fig. 9. Impact of interharmonic frequency.

Analytical sensitivity study is conducted to determine the accuracy requirement on $f_{os}$ estimation as follows. It is reasonable to assume 2 seconds of phasor data is used by FFT to determine $f_{os}$. This implies the frequency resolution of FFT is $1/T_w$=0.5Hz. In other words, the error of $f_{os}$ is about 0.5Hz (=Δ). This error will lead to error on $f_{IH2}$ estimation:

$$\varepsilon = \left|\frac{(f_1+f_{os})-[f_1+(f_{os}\pm\Delta)]}{(f_1+f_{os})}\right| = \left|\frac{\Delta}{f_1+f_{os}}\right| < \frac{\Delta}{f_1} = 0.83\%$$

It can be seen that relative error of $f_{IH2}$ estimation is very small. Therefore, FFT method is sufficient. In practice, an oscillation event should last at least 5 seconds to be considered as an event. This means that $T_w$ is expected to be at least 5sec.

Studies have also been conducted for cases where the actual system frequency $f_1$ deviates from $f_0$. The impact is significant. Fortunately, this problem be addressed by using more advanced estimation algorithms. Due to space limitation, this subject is omitted.

### C. Summary

The main findings of this section are summarized below:

- There is no need to use sophisticated algorithms to determine $f_{os}$. Simple FFT analysis should be sufficient;
- The accuracy of interharmonic estimation is not sensitive to the presence of one or two interharmonics;
- The accuracy is mildly sensitive to the weighting function and the PMU sampling rate;
- The accuracy is very sensitive to 1) the timestamp of the PMU data, and 2) the derivation of actual frequency from $f_0$.
- If $f_{os}$>20Hz which implies $f_{IH2}$>80Hz, the phasor data is not adequate for interharmonic estimation (see Fig.9).

## V. Source Location App “OSLocator”

The proposed method has been implemented into a cloud-based APP for public test and use. All information about this app is self-explanatory. The app can be accessed from https://oslocator.vercel.app/. The app is hosted by a well-known cloud computing provider Vercel.

### A. Input and Output Data

The input phasor data is in an Excel file using the format adopted by the IEEE PES Task Force on Forced Oscillations. The TF’s website provides several oscillation events using this format [11]. The format is shown on Fig. 11. Note that the frequency data is not used by the App. The output data is also in an Excel file with four sheets. The first three sheets store voltage, current, power results respectively for $f_1$, IH1, and IH2 components. The fourth sheet records the parameters entered by the user when running the app.

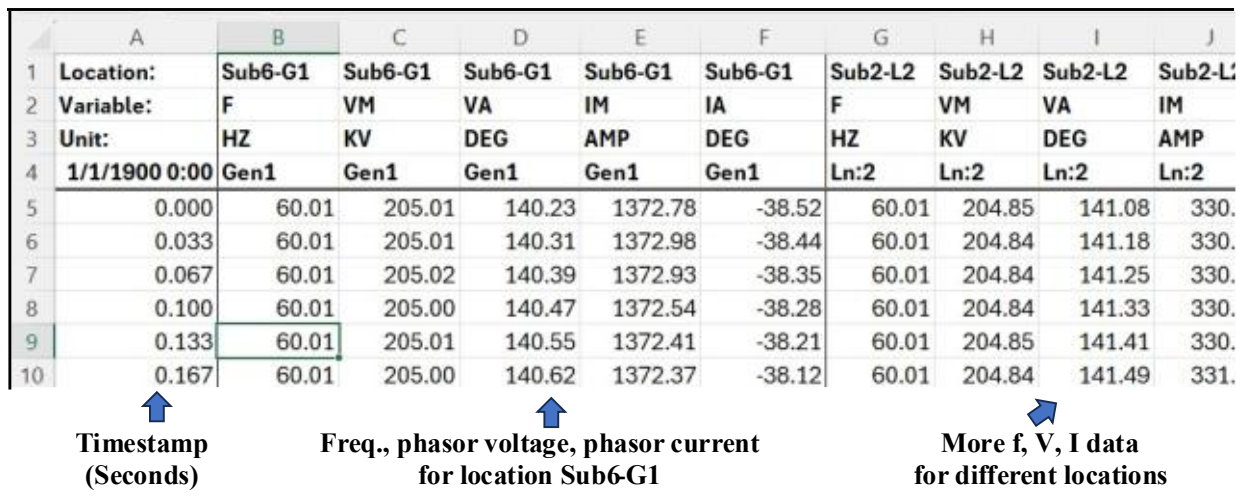

|  | A | B | C | D | E | F | G | H | I | J |
|---|---|---|---|---|---|---|---|---|---|---|
| 1 | Location: | Sub6-G1 | Sub6-G1 | Sub6-G1 | Sub6-G1 | Sub6-G1 | Sub2-L2 | Sub2-L2 | Sub2-L2 | Sub2-L |
| 2 | Variable: | F | VM | VA | IM | IA | F | VM | VA | IM |
| 3 | Unit: | HZ | KV | DEG | AMP | DEG | HZ | KV | DEG | AMP |
| 4 | 1/1/1900 0:00 | Gen1 | Gen1 | Gen1 | Gen1 | Gen1 | Ln:2 | Ln:2 | Ln:2 | Ln:2 |
| 5 | 0.000 | 60.01 | 205.01 | 140.23 | 1372.78 | -38.52 | 60.01 | 204.85 | 141.08 | 330. |
| 6 | 0.033 | 60.01 | 205.01 | 140.31 | 1372.98 | -38.44 | 60.01 | 204.84 | 141.18 | 330. |
| 7 | 0.067 | 60.01 | 205.02 | 140.39 | 1372.93 | -38.35 | 60.01 | 204.84 | 141.25 | 330. |
| 8 | 0.100 | 60.01 | 205.00 | 140.47 | 1372.54 | -38.28 | 60.01 | 204.84 | 141.33 | 330. |
| 9 | 0.133 | 60.01 | 205.01 | 140.55 | 1372.41 | -38.21 | 60.01 | 204.85 | 141.41 | 330. |
| 10 | 0.167 | 60.01 | 205.00 | 140.62 | 1372.37 | -38.12 | 60.01 | 204.84 | 141.49 | 331. |

Fig. 10. Sample input phasor data.

## B. *Application Example*

The OSLocator app is demonstrated here using ISO-NE-Case5 in [11]. The first screen, shown in Fig. 12, is to enter file location and parameters. The parameters include:

- Starting and ending time of the data segment to be used for analysis. The default is to use the entire data.
- The anchor of timestamp used by PMU. Default=50%
- The PMU sampling rate if known. Default=32.

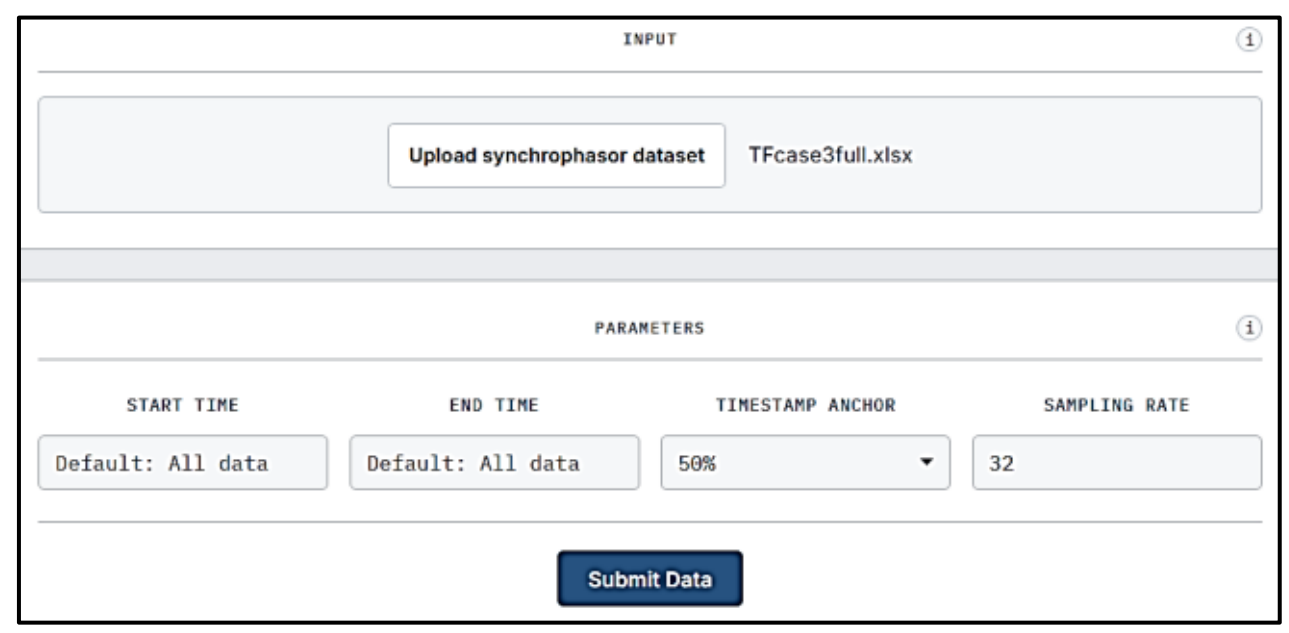

Fig. 11. Sample input phasor data.

The App then finds the location with the strongest oscillation. The voltage data is used to determine $f_{os}$. The results are displayed on the screen illustrated in Fig. 12.

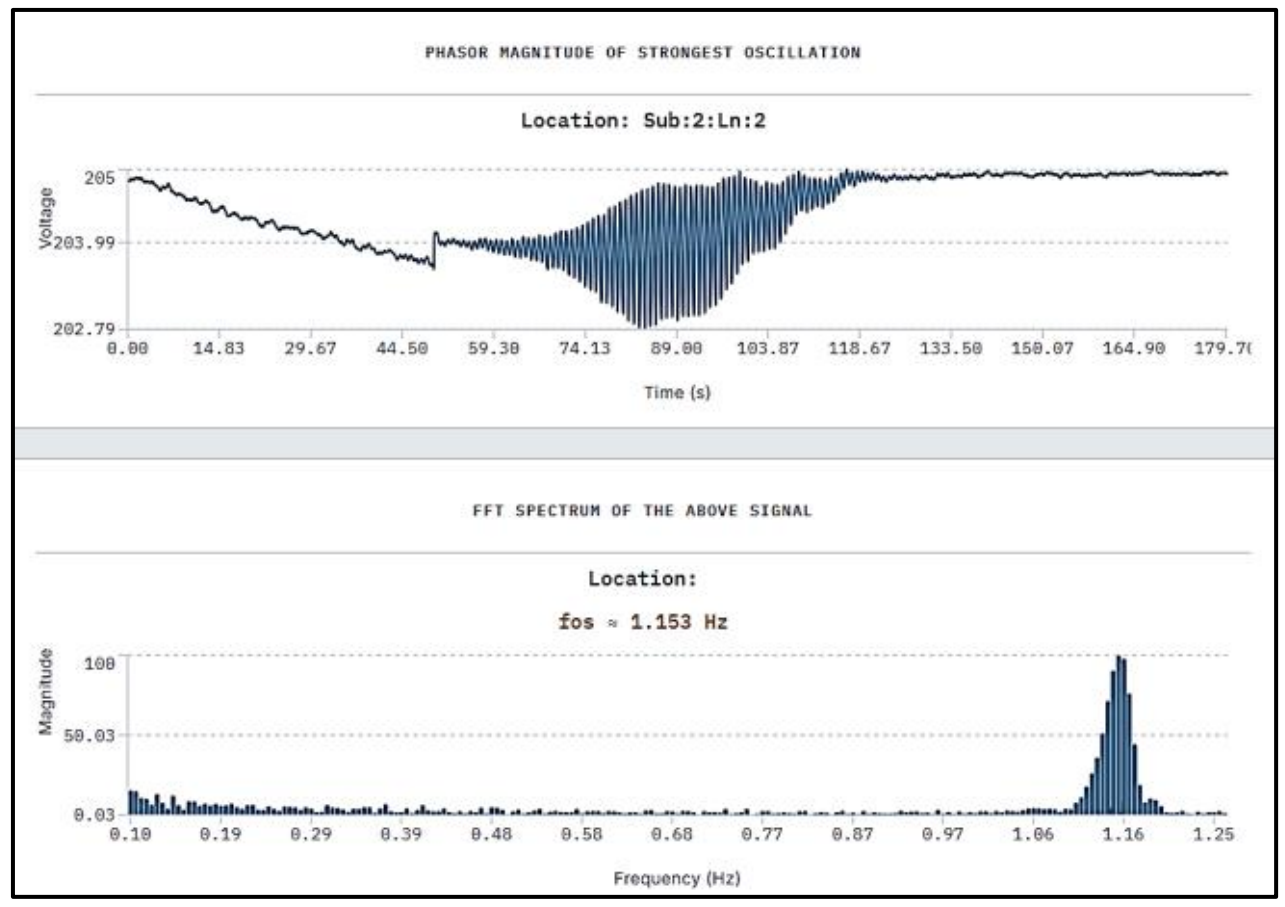

Fig. 12. Determination of oscillation frequency.

The final interharmonic results are saved in an Excel file for user to download. It is also displayed on screen as shown on Fig. 13. User can select different locations and variables to display. The timestamp anchor for the results is the timestamp of the first phasor data point of the oscillation periods used to extract the interharmonic results.

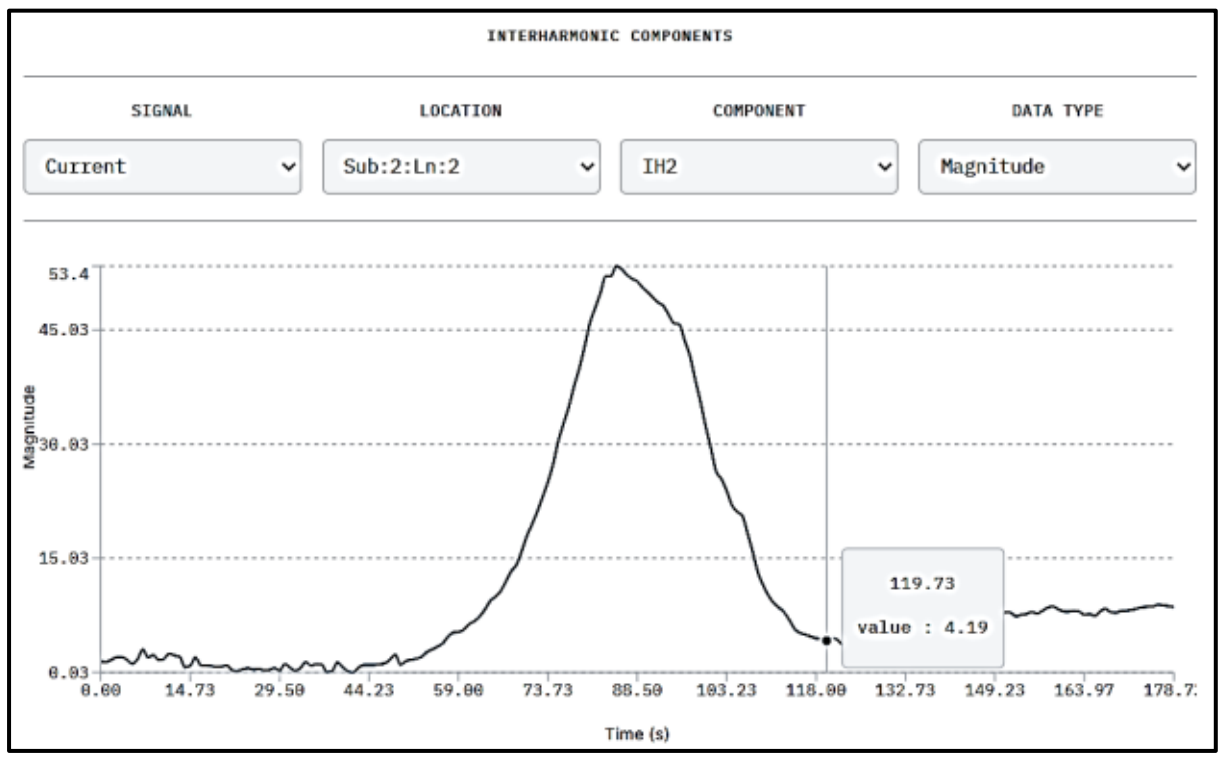

Fig. 13. Results displayed on screen.

# VI. Conclusions

This paper presented a method for locating power system oscillation sources by extracting interharmonic information from synchrophasor data. The method extends interharmonic-power-based source location to practical monitoring environments where waveform data are not routinely available.

Field-data tests and sensitivity studies indicate that PMU phasor data can support useful interharmonic extraction and source-location analysis under appropriate conditions. The developed OSLocator cloud application provides an initial platform for applying the method to field PMU data.